\documentstyle[12pt]{article}
\catcode`@=11
\newcount\@tempcntc
\def\@citex[#1]#2{\if@filesw\immediate\write\@auxout{\string\citation{#2}}\fi
  \@tempcnta\z@\@tempcntb\m@ne\def\@citea{}\@cite{\@for\@citeb:=#2\do
    {\@ifundefined
       {b@\@citeb}{\@citeo\@tempcntb\m@ne\@citea\def\@citea{,}{\bf ?}\@warning
       {Citation `\@citeb' on page \thepage \space undefined}}%
    {\setbox\z@\hbox{\global\@tempcntc0\csname b@\@citeb\endcsname\relax}%
     \ifnum\@tempcntc=\z@ \@citeo\@tempcntb\m@ne
       \@citea\def\@citea{,}\hbox{\csname b@\@citeb\endcsname}%
     \else
      \advance\@tempcntb\@ne
      \ifnum\@tempcntb=\@tempcntc
      \else\advance\@tempcntb\m@ne\@citeo
      \@tempcnta\@tempcntc\@tempcntb\@tempcntc\fi\fi}}\@citeo}{#1}}
\def\@citeo{\ifnum\@tempcnta>\@tempcntb\else\@citea\def\@citea{,}%
  \ifnum\@tempcnta=\@tempcntb\the\@tempcnta\else
   {\advance\@tempcnta\@ne\ifnum\@tempcnta=\@tempcntb \else \def\@citea{--}\fi
    \advance\@tempcnta\m@ne\the\@tempcnta\@citea\the\@tempcntb}\fi\fi}
\catcode`@=12

\def\theequation{\arabic{section}.\arabic{equation}}
\def\barr{\begin{array}}
\def\earr{\end{array}}
\def\beq{\begin{equation}}
\def\eeq{\end{equation}}
\def\bea{\begin{eqnarray}}
\def\eea{\end{eqnarray}}
\def\bmath{\begin{displaymath}}
\def\emath{\end{displaymath}}
\def\bq{\begin{quote}}
\def\eq{\end{quote}}
\def\Re{\mbox{Re}}

\def\cA{{\cal A}}
\def\cB{{\cal B}}
\def\cC{{\cal C}}
\def\cD{{\cal D}}
\def\cE{{\cal E}}
\def\cT{{\cal T}}
\def\veps{\varepsilon}
\def\slash#1{\setbox0=\hbox{$#1$}#1\hskip-\wd0\hbox to\wd0{\hss\sl/\/\hss}}

\def\fortp#1{{\em Fort.\ Phys.\ }{\bf #1}}

\def\npb#1{{\em Nucl.\ Phys.\ }{\bf B#1}}
\def\plb#1{{\em Phys.\ Lett.\ }{\bf A#1}}
\def\plb#1{{\em Phys.\ Lett.\ }{\bf B#1}}
\def\prl#1{{\em Phys.\ Rev.\ Lett.\ }{\bf #1}}

\def\prd#1{{\em Phys.\ Rev.\ }{\bf D#1}}

\def\zpc#1{{\em Z.\ Phys.\ }{\bf C#1}}
\begin{document}

\centerline{DESY 94-020\hfill ISSN 0418-9833}
\centerline{RAL/94-021\hfill}
\centerline{February 1994\hfill}
\vskip1cm
\begin{center}
{\bf{\Large QUANTUM EFFECTS ON HIGGS-BOSON}}\\[0.3cm]
{\bf{\Large PRODUCTION AND DECAY}}\\[0.3cm]
{\bf{\Large DUE TO MAJORANA NEUTRINOS}}\\[2cm]
{\large\sf B.A.~Kniehl}$^{a}$\hspace{0.13cm}
{\large\sf and\hspace{0.13cm} A.~Pilaftsis}$^{b}$
\footnote[1]{E-mail address: pilaftsis@vax2.rutherford.ac.uk}\\[0.4cm]
$^{a}$ {\em II. Institut f\"ur Theoretische Physik,
Universit\"at Hamburg, 22761 Hamburg, Germany}\\[0.3cm]
$^{b}$ {\em Rutherford Appleton Laboratory, Chilton, Didcot, Oxon, OX11 0QX,
England}
\end{center}
\vskip5cm
\centerline {\bf ABSTRACT}

We analyze the phenomenological implications for new electroweak physics
in the Higgs sector in the framework of $SU(2)_L\otimes U(1)_Y$ theories
that naturally predict heavy Majorana neutrinos.
We calculate the one-loop Majorana-neutrino contributions to the decay rates
of the Higgs boson into pairs of quarks and intermediate bosons
and to its production cross section via bremsstrahlung in $e^+e^-$ collisions.
It turns out that these are extremely small in three-generation models.
On the other hand, the sizeable quantum corrections generated by a conventional
fourth generation with a Dirac neutrino may be screened considerably in the
presence of a Majorana degree of freedom.

\newpage

\section{Introduction}
\indent

The well-established ``see-saw" mechanism, as suggested
by Yanagida, Gell-Mann, Ramond, and Slansky~\cite{YAN} in grand unified
theories~\cite{GUT}, could serve as
a natural solution to the problem of the smallness in mass of the three
known neutrinos, $\nu_e$, $\nu_\mu$, and $\nu_\tau$,
provided that these are Majorana particles.
Such a solution entails the existence of very heavy neutral leptons, which
also have to be of Majorana type. If heavy Majorana neutrinos are assumed to
be realized in nature at the mass scale of 1--10~TeV, they may manifest
themselves in lepton-number violating processes involving the $Z$ and Higgs
bosons~\cite{KPS,APetal} or through non-universality effects
in diagonal leptonic $Z$-boson decays~\cite{BKPS}. Their presence
may also influence~\cite{BS,Hansi} the electroweak oblique parameters,
$S$, $T$, $U$ (or $\veps_1$, $\veps_2$, $\veps_3$)~\cite{PT},
as well as the parameters $X$, $Y$, $Z$~\cite{XYZ} introduced recently.
The masses of such heavy Majorana neutrinos and their couplings to ordinary
matter should satisfy a large set of stringent constraints coming from a
global analysis of charged-current universality, neutral-current effects,
and other low-energy data~\cite{LL}, which set severe limits on the prospects
of observing these particles in high-energy experiments.

Making use of the full power of existing data to constrain new electroweak
physics, we find that, in a large class of extensions of the minimal Standard
Model (SM) by Majorana neutrinos, the Higgs sector is feebly confined.
We demonstrate this by elaborating minimal scenarios which extend the
field content of the SM by introducing right-handed neutrinos~\cite{YAN,ZPC}.
In addition, new electroweak physics may arise from the possible existence of
fourth-generation Majorana neutrinos~\cite{HP}, which can be added in a
natural way without conflicting with the data from the CERN Large Electron
Positron Collider (LEP) and the SLAC Linear Collider (SLC)~\cite{HP,DR}.
Such scenarios can
account also for the mass hierarchy problem of the light neutrinos,
since the light neutrinos acquire their masses radiatively at the two-loop
level~\cite{BM}.
By the same token, this resolves the solar-neutrino
deficit problem~\cite{solar}, through the Mikheyev-Smirnov-Wolfenstein (MSW)
mechanism~\cite{MSW}. However, it has recently been argued
that the Hill-Paschos model~\cite{HP} containing
Majoron fields ($J$) violates astrophysical constraints by
predicting too big $Jee$, $Juu$, and $Jdd$ couplings~\cite{MZ}.
Subsequently, it has been shown that, even in Majoron models with {\em three}
generations,
astrophysical constraints limit considerably the testability of such
models by terrestrial experiments~\cite{APmaj}.
In order to avoid that our analysis depends
on whether Majoron scalars are present in the model or not,
we shall consider Majoronless models by assuming that
the Majorana mass terms in the Yukawa sector are bare. Similar Majoronless
scenarios can be realized if the SM gauge group is extended by an additional
hypercharge group, $U(1)_{Y'}$~\cite{BCK}. In such theories, Majoron
fields are completely absent. For simplicity, we shall also assume the absence
of Majoron-triplet scalars~\cite{GR}, as they seem to be ruled out
by the present LEP data.

This work is organized as follows: In Sect.~2, we shall give a short
description
of the basic low-energy structure of the SM with right-handed neutrinos.
In Sect.~3, we shall compute the quantum corrections to the
$q\bar{q}H$, $WWH$, $ZZH$, and $ZAH$ vertices induced by heavy Majorana
neutrinos along with additional charged leptons and quarks.
We shall also briefly outline our renormalization scheme.
In Sect.~4, we shall discuss our numerical results and
assess the possibility of discovering in the Higgs sector new electroweak
physics beyond the minimal SM.
Section~5 contains our conclusions.

\section{{\boldmath $SU(2)_L\otimes U(1)_Y$ } theories with right-handed
\newline neutrinos}
\indent

Heavy Majorana neutrinos can naturally be predicted in extensions of the SM
where $\Delta L=2$ operators have been introduced in the Yukawa sector by the
inclusion of right-handed neutrinos. The quark sector of
such extensions is similar to that of the minimal SM.
For the leptonic sector, we shall adopt the conventions of Ref.~\cite{ZPC}.
Specifically, the interactions of the Majorana neutrinos, $n_i$, and charged
leptons, $l_i$, with the gauge bosons, $W^\pm$ and $Z$, and the Higgs boson,
$H$, are described by the following Lagrangians~\cite{ZPC}:
\bea
{\cal L}_{int}^W &=& -\ \frac{g_W}{2\sqrt{2}} W^{-\mu}\
\sum_{i=1}^{n_G}\sum_{j=1}^{2n_G}
{\bar{l}}_i \ B_{l_ij} {\gamma}_{\mu} (1-{\gamma}_5) \ n_j \ + \
\mbox{h.c.}\ ,
\\[0.3cm]
{\cal L}_{int}^Z &=& -\ \frac{g_W}{4\cos\theta_W}  Z^\mu\
\sum_{i,j=1}^{2n_G}
\bar{n}_i \gamma_\mu \Big[ i\mbox{Im}C_{ij}\ -\ \gamma_5\mbox{Re}C_{ij}
\Big] n_j\ ,\\[0.3cm]
{\cal L}^H_{int} &=& -\ \frac{g_W}{4M_W}\ H\
\sum_{i,j=1}^{2n_G}
\bar{n}_i \Bigg[ (m_i+m_j)\mbox{Re}C_{ij}
+\ i\gamma_5 (m_j-m_i)\mbox{Im}C_{ij} \Bigg] n_j\ ,
\eea 
where $m_i$ are the masses of $n_i$ and
$B$ and $C$ are $n_G\times 2n_G$ and $2n_G\times 2n_G$ dimensional
mixing matrices, respectively, with $n_G$ being the number of generations.
These matrices are defined as
\bea
B_{l_ij}\ &=& \sum\limits_{k=1}^{n_G} V^l_{l_ik} U^{\nu\ast}_{kj}\ ,\\
C_{ij}\ &=&\ \sum\limits_{k=1}^{n_G}\ U^\nu_{ki}U^{\nu\ast}_{kj}\ ,
\eea 
where $V^l$ and $U^\nu$ are the leptonic Cabbibo-Kobayashi-Maskawa
(CKM) matrix and the unitary matrix that diagonalizes the $2n_G
\times 2n_G$ ``see-saw" neutrino mass matrix, respectively.
$B$ and $C$ satisfy a number of identities, which
guarantee the renormalizabilty of our minimally extended model,
namely~\cite{APetal,ZPC}
\bea
\sum\limits_{k=1}^{2n_G} B_{l_1k}B_{l_2k}^{\ast} =  {\delta}_{l_1l_2},
\qquad
\sum\limits_{k=1}^{2n_G} C_{ik}C^\ast_{jk} =  C_{ij}, \qquad
\sum\limits_{k=1}^{2n_G} B_{lk}C_{ki}  =   B_{li}, \qquad
\sum\limits_{k=1}^{n_G} B_{l_ki}^{\ast}B_{l_kj}  =  C_{ij},\\
\sum\limits_{k=1}^{2n_G} m_k C_{ik}C_{jk} =  0,\qquad
\sum\limits_{k=1}^{2n_G} m_k B_{lk}C^\ast_{ki} =  0,\qquad
\sum\limits_{k=1}^{2n_G} m_k B_{l_1k}B_{l_2k} =  0.\qquad\mbox{}
\eea
Equations~(2.6) and (2.7) allow us to verify the cancellations of ultraviolet
divergences in the one-loop renormalizations of the $q\bar{q}H$, $WWH$, $ZZH$,
and $ZAH$ vertices.
The only information used to obtain all these identities is the
gauge structure of the SM. Therefore, our theoretical analysis does
not depend on the explicit form of the Majorana-neutrino mass matrix.
In fact, a vast number of possible mass ans\"atze have been proposed
in the literature during the last few years~\cite{ansatz}. However, all these
mass models possess the very same low-energy gauge structure discussed here
and thus obey the sum rules of Eqs.~(2.6) and~(2.7). In other
words, the specific form of any such mass matrix produces only supplementary
relations between $m_i$, $B_{li}$, and $C_{ij}$ on top of the identities of
Eqs~(2.6) and~(2.7).
Similarly, the quark mass matrices given, {\it e.g.}, by the Fritzsch texture
lead to additional relations between the quark masses
and CKM mixings~\cite{FR}.

Another interesting feature of the SM with right-handed neutrinos
is that out-of-equilibrium
lepton-number-violating decays of heavy Majorana neutrinos can generate a
non-zero lepton number
($L$)~\cite{FT} in the universe through the $L$-violating interactions
of Eqs.~(2.1)--(2.3). This excess in $L$ then gives rise to a baryon-number
($B$) asymmetry of the universe via the $(B+L)$-violating sphaleron
interactions, which are in thermal equilibrium above the critical temperature
of the electroweak phase transition~\cite{KRS}. This mechanism has received
much
attention recently, and many studies have been devoted to
constrain the $(B-L)$-violating mass scale and so to derive a lower
mass bound for the heavy Majorana neutrinos~\cite{MAL,BY,CDEO,HDGR,CKO},
exploiting the dramatic effect of out-of-equilibrium conditions for the
$\Delta L=2$ operators.
Yet, it was conceivable that certain scenarios predicting
heavy Majorana neutrinos with $m_N=1$--10~TeV could naturally account for the
existing $B$ asymmetry in the universe (BAU)~\cite{MAL}.
Furthermore, on the basis of a two-generation scenario of right-handed
neutrinos, it was argued~\cite{BY} that the $m_N$ lower bound of
$\sim 1$~TeV had been underestimated by many orders of magnitude.
As a consequence, the opportunity of probing Majorana-neutrino physics
at LEP energies would be extinguished practically.
Fortunately, very recently, careful inspection of chemical potentials
in the framework of three generations with lepton-flavour mixings have
revealed that all these stringent bounds can be weakened dramatically
and are quite model dependent~\cite{HDGR,CKO}. In particular,
it is sufficient that one individual lepton number, $L_e$ say,
is conserved in order to protect any primordial BAU generated at the
$GUT$ scale from being erased by sphalerons, even if operators with
$\Delta L_{l_i}\neq \Delta L_e$ were in thermal equilibrium~\cite{HDGR}.
The reason is that
sphalerons generally conserve the quantum numbers
$B/3-L_{l_i}$~\cite{CDEO,HDGR}
and thus protect any pre-existing BAU from being washed out.
Similar conclusions have been drawn in Ref.~\cite{CKO}, where it was suggested
that the BAU may be preserved by an asymmetry in the number of right-handed
electrons.
These new observations support our assumptions concerning viable scenarios
of heavy Majorana neutrinos with masses in the TeV range, which
couple very feebly to a separate leptonic quantum number, so that,
for instance, $\Delta L_e =0$.

\setcounter{equation}{0}
\section{Heavy Majorana neutrinos and Higgs phenomeno-
\newline logy}
\indent

In this section, we shall analyze quantitatively the implications
of Majorana neutrinos for the Higgs sector at the quantum level.
Specifically, we shall consider the Higgs-boson decays $H\to q\bar{q}$,
$H\to WW$, and $H\to ZZ$ as well as the production mechanism
$e^+e^-\to ZH$.
Since all these processes have been studied at one loop in the SM
already~\cite{Bernd}, we shall focus attention on the quantum corrections
induced by the extended fermion sector described in Sect.~2.
Each generation contains two Majorana neutrinos, one charged lepton,
an up-type quark, and a down-type quark, so that the anomalies cancel.
Since we wish to estimate the size of new physics both in
three- and four-generation models,
we shall keep the number of generations arbitrary.

Loop calculations in electroweak physics are frequently carried out in the
on-mass-shell scheme, which uses the fine-structure constant, $\alpha$,
and the physical particle masses as basic parameters~\cite{bhs}.
One drawback of this scheme is the occurrence of large logarithms
connected with light charged fermions, which drive the renormalization scale
of $\alpha$ from $m_e$ to $M_Z$ and artificially enhance the corrections.
These logarithms may be removed from the corrections and resummed by
expressing the Born result in terms of the Fermi constant,
$G_F=\left[\pi\alpha/\sqrt2s_w^2M_W^2(1-\Delta r)\right]$,
where $\Delta r$ embodies the non-photonic corrections to the muon
decay rate~\cite{delr}.
As a consequence, a multiple of $\Delta r$ is added to the correction
in such a way that the large logarithms are exactly cancelled.
This procedure is sometimes called modified on-mass-shell (MOMS) scheme.

\subsection{The decay {\boldmath $H \to q\bar{q}$}}
\indent

The one-loop electroweak corrections to the $H\to q\bar{q}$ decay width
are well known within the minimal SM~\cite{Hqq}.
The contribution due to the fermion sector modified by the presence
of Majorana neutrinos as described in Sect.~2, relative to the
Born decay width,
\beq
\Gamma_0(H\to q\bar{q})\ =\ \frac{N_cG_FM_Hm_q^2}{4\pi\sqrt{2}}
\left(\ 1\ -\ \frac{4m^2_q}{M^2_H}\ \right)^{3/2},
\eeq 
with $N_c=3$, may be calculated in the MOMS scheme from
\beq
\delta_q\ =\ -\ \Re \Pi_{HH}'(M^2_H)\ -\ \frac{\Pi_{WW}(0)}{M^2_W}
\ -\ \frac{2}{s_wc_w}\ \frac{\Pi_{ZA}(0)}{M^2_Z}\ -\ \Delta r_{direct},
\eeq 
where $\Pi_{WW}$, $\Pi_{ZA}$, and $\Pi_{HH}$ denote unrenormalized
vacuum-polarization functions and $\Delta r_{direct}$ comprises the
non-photonic vertex and box contributions to $\Delta r$~\cite{delr}.
As a consequence of electromagnetic gauge invariance, $\Pi_{ZA}(0)=0$
for fermionic contributions.
Throughout this work, we shall assume that the novel heavy Majorana neutrinos
couple so weakly to the electron and muon that their contribution to
$\Delta r_{direct}$ may be neglected, which agrees with observations
by other authors~\cite{LL,BCK}.

Note that Majorana neutrinos do not contribute through triangle diagrams
to $\Gamma(H\to q\bar q)$ at one loop.
This is quite different for lepton pair production.
However, the decays into the known lepton flavours are suppressed by the
smallness of the Yukawa couplings,
and this is not expected to be changed by virtual Majorana-neutrino effects.
We shall leave the study of the leptonic decays for future work.
Furthermore, it is interesting to observe that the $Z$ and Higgs bosons can
mix via loops involving Majorana neutrinos.
Such amplitudes, which do not exist in the SM, render the branching
ratios of the $t_L\bar t_L$ and $t_R\bar t_R$ channels,
where $L$ and $R$ label the helicity states, different,
which is a signal for CP violation~\cite{IKP}.
However, these contributions cancel when the helicities are summed over.

The fermionic contributions to the vacuum polarizations may be
cast in the general forms,
\bea
\Pi_{WW}(q^2)&=& \frac{G_FM^2_W}{\sqrt{2}}\Bigg[ |B_{li}|^2 \left(
\Pi_V(q^2,m_i,m_l)+\Pi_V(q^2,m_i,-m_l)\right) \nonumber\\
&&+\ N_c|V_{ud}|^2\left( \Pi_V(q^2,m_u,m_d)+\Pi_V(q^2,m_u,-m_d)\right)\Bigg],\\
\Pi_{ZZ}(q^2)&=&\frac{G_FM^2_Z}{2\sqrt{2}}\Bigg[|C_{ij}|^2
\left(\Pi_V(q^2,m_i,m_j)+\Pi_V(q^2,m_i,-m_j)\right)\nonumber\\
&&-\ \Re C^2_{ij}
\left(\Pi_V(q^2,m_i,m_j)-\Pi_V(q^2,m_i,-m_j)\right)\nonumber\\
&&+\ v^2_l\Pi_V(q^2,m_l,m_l)+\Pi_V(q^2,m_l,-m_l)\nonumber\\
&&+\ N_c\left( v^2_q\Pi_V(q^2,m_q,m_q)+\Pi_V(q^2,m_q,-m_q) \right)\Bigg],\\
\Pi_{ZA}(q^2)&=& -\sqrt{2}G_FM^2_Zs_wc_w\Big[ -v_l\Pi_V(q^2,m_l,m_l)\ +\
N_cv_qQ_q\Pi_V(q^2,m_q,m_q)\Big],\\
\Pi_{HH}(q^2) &=& \frac{G_F}{2\sqrt{2}} \Bigg[\hspace{-1.7pt}
\left( |C_{ij}|^2 (m^2_i + m^2_j)+2m_im_j \Re C_{ij}^2 \right)\hspace{-1.7pt}
\left(\Pi_S(q^2,m_i,m_j) + \Pi_S(q^2,m_i,-m_j) \right) \nonumber\\
&&+\ \left( 2m_im_j|C_{ij}|^2+(m^2_i+m^2_j)\Re C_{ij}^2 \right)
\left(\Pi_S(q^2,m_i,m_j) - \Pi_S(q^2,m_i,-m_j) \right) \nonumber\\
&&+\ 4m^2_l\Pi_S(q^2,m_l,m_l)\ +\ 4N_cm^2_q\Pi_S(q^2,m_q,m_q)\Bigg],
\eea 
where $v_f=2T_f-4s_w^2Q_f$ is the $Zf\bar f$ vector coupling, $T_f$ is the
weak isospin of $f$, $Q_f$ is its electric charge in units of the positron
charge, $V_{ud}$ is the $n_G\times n_G$ CKM matrix of the quark sector,
and the scalar and vector functions, $\Pi_S$ and $\Pi_V$, are listed in the
Appendix.
Here and in the following, it is understood that indices are to be summed
over when they appear more than once in an expression.
For later use, we have presented also $\Pi_{ZZ}$ and $\Pi_{ZA}$.
We postpone the numerical discussion of the new virtual effects to
Sect.~4.

\subsection{The decay {\boldmath $H \to VV$}}
\indent

The one-loop renormalization of the $H\to WW$ and $H\to ZZ$ decay widths
in the minimal SM is described in Refs.~\cite{HWW,HZZ}.
Modifications of the fermion sector affect these decay widths through the
$WWH$ and $ZZH$ triangle diagrams depicted in Figs.~1(a) and (b),
respectively.
Assigning incoming four-momenta and Lorentz indices, $(p,\mu)$ and $(k,\nu)$,
to the vector bosons, $V$, the renormalized vertex function takes
the form
\bea
\cT_{VVH}^{\mu\nu}&=& 2^{5/4}G_F^{1/2}M^2_V\ \Bigg[D_{VVH}(a,b,c)\
k^\mu p^\nu\ +\ \left(1+\hat{E}_{VVH}(a,b,c)\right)\ \mbox{g}^{\mu\nu}
\nonumber\\
&&+\ iF_{VVH}(a,b,c)\ \varepsilon^{\mu\nu\rho\sigma}p_\rho k_\sigma \Bigg],
\eea 
where $a=p^2$, $b=k^2$, $c=(p+k)^2$, and we have discarded terms with
$p^\mu$ or $k^\nu$ anticipating that, in our applications, the vector bosons
are real or couple to conserved currents.
The hatted quantity has been renormalized by including its counterterm,
\beq
\hat{E}_{VVH}(a,b,c)\ =\ E_{VVH}(a,b,c)\ +\ \delta E_{VVH}.
\eeq 
In the MOMS scheme, one has
\bea
\delta E_{VVH} & = & \Re\Bigg( \frac{\Pi_{VV}(M^2_V)}{M^2_V}\ -
\Pi_{VV}'(M^2_V) \Bigg)\ -\ \frac{1}{2}\ \Bigg(
\Re\Pi_{HH}'(M^2_H)\ +\ \frac{\Pi_{WW}(0)}{M^2_W} \nonumber\\
& & +\ \Delta r_{direct} \Bigg) \ -\ \frac{1}{s_wc_w}\
\frac{\Pi_{ZA}(0)}{M^2_Z}.
\eea 

In the SM, $F_{WWH}(a,a,c)=F_{ZZH}(a,b,c)=0$ for $a,b,c$ arbitrary,
due to CP conservation~\cite{HWW,HZZ}.
In the presence of Majorana neutrinos, $F_{VVH}(a,a,c)$ does not vanish,
in general, so that CP-violating interactions are generated.
If the vector-boson polarizations can be determined experimentally,
it is possible to construct asymmetries that measure the degree of
CP-nonconservation~\cite{IKP}.
However, the $F_{VVH}$ term drops out when we sum over all vector-boson
polarizations, which we shall do to obtain the total $H\to VV$ decay rates.

In the considered class of models, the most general representations for the
$D_{VVH}$ form factors read
\bea
D_{WWH} &=& -\ \frac{G_F}{4\sqrt{2}} \Bigg[ B_{li}^\ast C_{ij}^\ast B_{lj}
\left( m_i \overline{\cD}(m_j,m_l,m_i)\ +\ m_j\overline{\cD}(m_i,m_l,m_j)
\right)\nonumber\\
&&+\ B^\ast_{li} C_{ij} B_{lj} \left(
m_i \overline{\cD}(m_i,m_l,m_j)\ +\ m_j\overline{\cD}(m_j,m_l,m_i)
\right)\nonumber\\
&&+\ 2m_l|B_{li}|^2\left( \cD(m_l,m_i,m_l)\ +\ \cD(m_l,-m_i,m_l)
\right) \nonumber\\
&&+\ 2N_cm_u|V_{ud}|^2\left( \cD(m_u,m_d,m_u)\ +\ \cD(m_u,-m_d,m_u)
\right) \nonumber\\
&&+\ 2N_cm_d|V_{ud}|^2\left( \cD(m_d,m_u,m_d)\ +\ \cD(m_d,-m_u,m_d)
\right) \Bigg],\\
D_{ZZH} &=& - \frac{G_F}{4\sqrt{2}} \Bigg[ \Re (C_{ik}C_{kj}C_{ji})
\left( m_i \overline{\cD}(m_k,m_j,m_i)\ +\ m_k\overline{\cD}(m_i,m_j,m_k)
\right)\nonumber\\
&&+\ \Re (C^\ast_{ik} C_{kj} C_{ji}) \left(
m_i \overline{\cD}(m_i,m_j,m_k)\ +\ m_k\overline{\cD}(m_k,m_j,m_i)
\right)\nonumber\\
&&-\ \Re (C_{ik} C_{kj}^\ast C_{ji})
\cD_-(m_i,m_j,m_k)\ -\
\Re (C_{ik} C_{kj} C_{ji}^\ast) \cD_+(m_i,m_j,m_k) \nonumber\\
&&+\ 2m_l\left( v_l^2\cD(m_l,m_l,m_l)\ +\ \cD(m_l,-m_l,m_l)
\right)\nonumber\\
&&+\ 2N_cm_q\left( v_q^2\cD(m_q,m_q,m_q)\ +\ \cD(m_q,-m_q,m_q)
\right) \Bigg],\\
D_{ZAH}&=&\sqrt{2}G_Fs_wc_w \Big[ -m_lv_l\cD(m_l,m_l,m_l)
+ N_cm_qv_qQ_q\cD(m_q,m_q,m_q) \Big],
\eea 
where we have suppressed the labels $a,b,c$ on both sides of the
equations.
The auxiliary functions $\cD$, $\overline{\cD}$, and $\cD_\pm$
are listed in the Appendix.
For later use, we have also presented the $D_{ZAH}$ form factor
appropriate to the $Z$-photon-Higgs vertex shown in Fig.~1(b).
Here $b$ is the photon invariant mass squared.
The expressions for $E_{WWH}$, $E_{ZZH}$, and $E_{ZAH}$ are similar to
Eqs.~(3.10)--(3.12), and the corresponding functions $\cE$, $\overline{\cE}$,
and $\cE_\pm$ are given in the Appendix.

Defining $r_V=\left(M^2_H/4M^2_V\right)$, $n_W=1$, and $n_Z=2$,
the Born approximation for the $H\to VV$ decay rate reads
\beq
\Gamma_0(H\to VV)\ =\ \frac{1}{n_V}\ \frac{G_FM^3_H}{8\pi\sqrt{2}}
\sqrt{1-\frac{1}{r_V}}\ \left(\ 1\ -\ \frac{1}{r_V}\ +\
\frac{3}{4r^2_V}\ \right).
\eeq 
The fermion sector extended by Majorana neutrinos
induces a relative correction to Eq.~(3.13),
which is given by
\beq
\delta_V = 2\Re\hat{E}_{VVH}(M_V^2,M_V^2,M_H^2) +
\frac{(1-1/r_V)[1-(1/2r_V)]}{1-1/r_V+(3/4r_V^2)}
M^2_H \Re D_{VVH}(M_V^2,M_V^2,M_H^2).
\eeq 
In Sect.~4, we shall evaluate this expression numerically.

\subsection{The reaction {\boldmath $e^+e^-\to ZH$}}
\indent

At LEP200 and future $e^+e^-$ linear colliders with $\sqrt s\le500$~GeV,
the bremsstrahlung process, $e^+e^-\to ZH$, will be the most copious source
of Higgs bosons in the intermediate mass range~\cite{BAK},
and it is important to have the radiative corrections to its cross section
well under control.
These have been calculated in the SM~\cite{EEZH} and in its minimal
supersymmetric extension~\cite{ralf}.
Here, we shall study the influence of virtual heavy Majorana neutrinos.

To start with, we consider the angular distribution,
which, at tree level, is given by
\beq
\frac{d\sigma_{ZH}^0}{d\cos\theta}\ =\ \frac{G^2_FM^6_Z
\sqrt{\lambda}}{16\pi s(s-M^2_Z)^2}\ (1+v^2_e)
\left( \ 1\ +\ \frac{\lambda}{8sM^2_Z}\ \sin\theta\ \right),
\eeq 
where $\theta$ is the angle defined by the electron and $Z$-boson
three-momenta in the centre-of-mass frame and
$\lambda=(s-M_Z^2-M_H^2)^2-4M_Z^2M_H^2$.
The corrections due to the fermion sector with Majorana neutrinos
described in Sect.~2 may be included by multiplying Eq.~(3.15) with
$(1+2\Re\delta_{ZH})$, where
\bea
\delta_{ZH}\hspace{-2.55pt}&=&\hspace{-2.55pt}
\hat E_{ZZH}(M_Z^2,s,M_H^2)+F(\theta)D_{ZZH}(M_Z^2,s,M_H^2)
+\frac{s}{s-M^2_Z} \left(\frac{\Re\Pi_{ZZ}(M^2_Z)}{M^2_Z}
-\frac{\Pi_{ZZ}(s)}{s}\right)\nonumber\\
&&+\ \frac{1}{2}\Re\left(\frac{\Pi_{ZZ}(M^2_Z)}{M^2_Z}-\Pi_{ZZ}'(M^2_Z)\right)
\ -\ \frac{\Pi_{WW}(0)}{M^2_W}\ -\ \frac{1}{2}\Re\Pi_{HH}'(M^2_H)\nonumber\\
&&+\ \frac{4s_wc_wv_e}{1+v_e^2} \Bigg[\frac{s-M^2_Z}{s}\Big(
E_{ZAH}(M_Z^2,s,M_H^2)\ +\ F(\theta)D_{ZAH}(M_Z^2,s,M_H^2)\Big)\ -\
\frac{\Pi_{ZA}(s)}{s}\nonumber\\
&&+\ \frac{c_w}{s_w}\Re\left(
\frac{\Pi_{ZZ}(M^2_Z)}{M^2_Z}\ -\ \frac{\Pi_{WW}(M^2_W)}{M^2_W}\right)\ \Bigg],
\eea 
Here, all angular dependence is carried by
\beq
F(\theta)\ =\ \frac{(M^2_H-M^2_Z-s)\lambda\sin^2\theta}
{2( 8sM^2_Z\ +\ \lambda\sin^2\theta)}.
\eeq 
As before, we have assumed that the couplings of the electron to the
heavy Majorana neutrinos are suppressed~\cite{LL},
so that $e^+e^-H$ triangle and $e^+e^-ZH$ box contributions are shifted
from their SM values by insignificant amounts,
which may safely be neglected.

As for the integrated cross section, the Born result is
\beq
\sigma_{ZH}^0\ =\ \frac{G_F^2M^6_Z\sqrt{\lambda}}{8\pi s
(s-M^2_Z)^2}(1+v_e^2)\left(\ 1\ +\ \frac{\lambda}{12sM^2_Z}
\ \right),
\eeq
and the correction factor is
\beq
\left(1+2\Re\delta_{ZH}\Big|_{\sin^2\theta =2/3}\right).
\eeq 
The phenomenological implications of these results will be examined
in the next section.

\setcounter{equation}{0}
\section{Numerical results and discussion}
\indent

In Sect.~3, we have collected the analytic results for the one-loop
corrections to the rates of Higgs-boson production via $e^+e^-\to ZH$
and its decays to $q\bar q$, $W^+W^-$, and $ZZ$ pairs
in the context of three- and four-generation models with Majorana neutrinos.
We are now in a position to explore the phenomenological consequences
of our results.

To start with, we consider extensions of the SM by three right-handed
neutrinos.
We find that the relative corrections to the Higgs-boson observables under
consideration are shifted from their SM values by very small amounts,
which are typically of the order of a few tenths of a percent.
Similar observations have been made in connection with the oblique
parameters $S$, $T$, and $U$~\cite{Hansi}.

In the following, we shall thus concentrate on models that naturally
accommodate a fourth generation with Majorana neutrinos, adopting the
scenario proposed by Hill and Paschos~\cite{HP}.
For reasons mentioned in the Introduction, we take the Majorana masses
appearing in the Lagrangian to be bare.
Specifically, the fourth generation consists of two Majorana neutrinos,
$N_1$ and $N_2$, one charged lepton, $E$, one up-type quark, $t'$,
and one down-type quark, $b'$.
We assume that $E$, $t'$, and $b'$ have SM couplings.
All these new particles must have masses in excess of $M_Z/2$
so as to escape detection at the LEP/SLC experiments.
The Majorana and Dirac masses of the Majorana system are related to the
physical masses by
$M=m_{N_2}-m_{N_1}$ and $m_D=\sqrt{m_{N_1}m_{N_2}}$,
respectively.
Conversely, one has $m_{N_{1,2}}=\sqrt{m_D^2+M^2/4}\mp M/2$.

Since global analyses suggest that the mixings between the new fermions
and the established ones are greatly suppressed~\cite{LL},
we neglect these couplings altogether in our analysis.
Our new-physics scenario thus effectively reduces to a one-generation model.
The interactions between the novel fermions and the weak bosons are
characterized by Eqs.~(2.1)--(2.3) with $n_G=1$.
The mixing matrices may be determined from the identities of
Eqs.~(2.6) and (2.7) with the result that
\bea
C_{N_1N_1}=\frac{m_{N_2}}{m_{N_1}+m_{N_2}},\quad
C_{N_2N_2}=\frac{m_{N_1}}{m_{N_1}+m_{N_2}},\quad
C_{N_1N_2}=-C_{N_2N_1}=i\frac{\sqrt{m_{N_1}m_{N_2}}}{m_{N_1}+m_{N_2}},\\
B_{EN_1}=\sqrt{\frac{m_{N_2}}{m_{N_1}+m_{N_2}}},\quad
B_{EN_2}=i\sqrt{\frac{m_{N_1}}{m_{N_1}+m_{N_2}}}.\qquad\qquad\qquad\qquad
\eea 
Inspired by the fact that the third-generation quarks participate only feebly
in the CKM mixing, we ignore the possibility of mixing of $t'$ and $b'$
with quark flavours of the first three generations, {\it i.e.}, we put
$V_{t'b'}=1$ and $V_{t'd}=V_{ub'}=0$ otherwise.

Our final results are displayed in Figs.~2--10.
Figures~2--4 refer to $H\to t\bar t$ [cf.\ Eq.~(3.2)],
Figs.~5--7 to $H\to W^+W^-$ [cf.\ Eq.~(3.14) for $V=W$],
and Figs.~8--10 to $e^+e^-\to ZH$ [cf.\ Eq.~(3.16) multiplied by two].
Our results for $H\to ZZ$ are very similar to those for $H\to W^+W^-$.
In fact, $\delta_Z$ differs from $\delta_W$ by less than 0.1\%
in the considered parameter range.
This may be understood by observing that, in our analysis, the mass scale of
new physics is much larger than the mass difference of the $W$ and $Z$ bosons,
so that the custodial symmetry is in effect.
In each set of figures, the first two are devoted to a fourth-generation
scenario with two mass-degenerate Majorana neutrinos,
which are equivalent to one standard Dirac neutrino, {\it i.e.},
$m_D=m_{N_1}=m_{N_2}$ and $M=0$,
while the third figure deals with the genuine Majorana case, $M>0$, for which
$m_{N_1}<m_D<m_{N_2}$.
Since we are mainly interested in the Majorana system,
we assume $m_E=m_D$, which has been identified as a natural choice~\cite{HP},
and $m_{t'}=m_{b'}=400$~GeV in order to reduce the number of parameters to be
varied independently.
Figures 2 and 5 (3 and 6) examine the dependence of the radiative corrections
on $M_H$ ($m_D$) for selected values of $m_D$ ($M_H$).
In Figs.~4 and 7, the $m_{N_1}$ dependence is analyzed for $m_D=400$~GeV
and several values of $M_H$.
The spikes in Figs.~2--7 arise from threshold effects in the Higgs
wave-function renormalization and are an artifact of treating the Higgs boson
as an asymptotic state despite its limited lifetime.
They occur when $M_H=2m_i$, where $i=N_1,N_2,E,t',b'$.
The corrections remain finite at these points,
which may be understood from arguments based on dispersion
relations~\cite{HWW}.
The $H\to W^+W^-$ triangle diagrams have thresholds at the same points.

In Figs.~2 and 3, we see that a virtual heavy Dirac neutrino,
with $m_D\gg M_H/2$,
produces a positive correction to $\Gamma\left(H\to t\bar t\right)$,
which increases with $M_H$ decreasing and/or $m_D$ increasing.
This agrees with previous observations made in connection with additional
heavy-fermion doublets~\cite{Hqq}.
{}From Fig.~4, we learn that this conventional heavy-flavour effect may be
reduced by virtue of a mass splitting between $N_1$ and $N_2$,
{\it i.e.}, the possible Majorana nature of the lepton sector.
In the mass range considered, the maximum shift in $\delta_t$ with respect
to the Dirac case is $-5\%$ and occurs at $m_{N_1}=M_H/2$.
In other words, the influence of new heavy flavours may be screened
by the existence of a Majorana degree of freedom.
A screening effect of Majorana origin was encountered also in the case of
the $T$ parameter, which measures isospin breaking~\cite{BS,Hansi}.
In the present case, however, the heavy flavours generate large corrections
even if their Dirac masses are degenerate.

In the case of $H\to W^+W^-$, loop effects due to a Dirac neutrino with
$m_D\gg M_H/2$ reduce the decay rate by an amount that increases with $M_H$
and/or $m_D$; see Figs.~5 and 6.
Similar observations have been reported in Refs.~\cite{HWW,HZZ}.
Again, the magnitude of this effect may be decreased appreciably
by allowing for a nonvanishing Majorana mass, $m_{N_2}-m_{N_1}$; see Fig.~7.
In the mass range considered, the maximum shift in $\delta_W$ is 7\%.

In Fig.~8, the $M_H$ dependence of the shift in $\sigma(e^+e^-\to ZH)$
induced by a conventional fourth generation with
$m_{N_1}=m_{N_2}=m_E=m_{t'}=m_{b'}=400$~GeV is shown for LEP200 energy
and three $\sqrt s$ values appropriate to future $e^+e^-$ linacs.
For $M_H<700$~GeV, the corrections are negative, decrease in magnitude
with $\sqrt s$ increasing, and are practically independent of $M_H$.
The spikes at $M_H=800$~GeV are again due to threshold effects in the
Higgs wave-function renormalization.
In Figs.~9 and 10, we concentrate on Higgs-boson production at LEP200
and a 500-GeV linac, assuming $M_H=70$ and 200~GeV, respectively.
In Fig.~9, we study how the conventional fourth-generation correction
varies with the Dirac-neutrino mass, $m_D$.
In addition to the threshold effects related to the Higgs wave function,
there are new possible thresholds at $\sqrt s=2m_i\
(i=N_1,N_2,E,t',b')$ and $\sqrt s=m_{N_1}+m_{N_2}$ originating
from the $s$-channel cut through the $ZZH$ and $ZAH$ triangle diagrams.
The one at $\sqrt s=2m_{N_1}$ is visible in both cases considered in Fig.~9.
Leaving aside the threshold effects, the correction is negative and
its magnitude grows quadratically with $m_D$.
At $m_D=400$~GeV, it reaches $-4.3\%$.
When we now turn on the Majorana mass, we may reduce the effect down to
the level of $-1\%$ without affecting the invisible $Z$-boson width;
see Fig.~10.
Again, the impact of heavy flavours is screened in the presence of
genuine Majorana neutrinos.

\setcounter{equation}{0}
\section{Conclusions}
\indent

We have investigated the influence of virtual heavy Majorana neutrinos
on some of the most relevant processes involving the Higgs boson,
namely, its decays into pairs of quarks and intermediate bosons
as well as its production via bremsstrahlung in $e^+e^-$ collisions.
We found that the Standard Model predictions are changed insignificantly
when the Dirac neutrinos of the established three generations are split
into light and heavy Majorana neutrinos.
The situation is very different in the fourth-generation scenario proposed
by Hill and Paschos~\cite{HP}.
Here, the Majorana nature of the lepton sector manifests itself in a
screening of the typical heavy-flavour effects.
This feature is familiar from the electroweak parameter $T$~\cite{BS,Hansi},
which measures the breaking of isospin.
In contrast to $T$, however, the Higgs observables are sensitive to the
novel heavy flavours even if they are degenerate in mass.

\noindent
{\bf Acknowledgements.} We wish to thank Steven Abel and Wilfried Buchm\"uller
for stimulating discussions concerning cosmological constraints on models
with heavy Majorana neutrinos.

\setcounter{equation}{0}
\def\theequation{\Alph{section}.\arabic{equation}}
\begin{appendix}
\section{Appendix}
\indent

In this paper, we evaluate the loop amplitudes using dimensional
regularization along with the reduction algorithm of Ref.~\cite{PV}.
In contrast to Ref.~\cite{PV}, we use the Minkowskian metric,
$\mbox{g}^{\mu\nu}=\mbox{diag}(1,-1,\ldots,-1)$.

The scalar and vector two-point functions occurring in Eqs.~(3.3)--(3.6)
are defined as
\bea
\Pi_S(q^2,m_1,m_2) &=&-\frac{1}{16\pi^2}\int\ \frac{d^nl}{i\pi^2}
\ \mbox{tr}\left(\frac{1}{\slash{l}+\slash{q}-m_2}\ \frac{1}{\slash{l}-m_1}
\right)\nonumber\\
&=& \frac{1}{8\pi^2}\Bigg[ \left(q^2-(m_1+m_2)^2\right) B_0(q^2,m_1^2,
m^2_2)\ -\ m_1^2\left( 1+B_0(0,m_1^2,m^2_1) \right) \nonumber\\
&& - m^2_2\left(1+B_0(0,m_2^2,m^2_2)\right) \Bigg],\\
\Pi_V(q^2,m_1,m_2) &=& \frac{1}{n-1} \left(\mbox{g}_{\mu\nu}
-\frac{q_\mu q_\nu}{q^2}\right) \frac{1}{16\pi^2}\int\ \frac{d^nl}{i\pi^2}
\ \mbox{tr}\left(\gamma^\nu\, \frac{1}{\slash{l}+\slash{q}-m_2}
\ \gamma^\mu\, \frac{1}{\slash{l}-m_1}
\right)\nonumber\\
&=& \frac{1}{12\pi^2}\Bigg[ \left(q^2-\frac{m_1^2+m^2_2}{2}
+3m_1m_2-\frac{(m^2_1-m^2_2)^2}{2q^2}\right) B_0(q^2,m_1^2,m^2_2)\nonumber\\
&&+\ m_1^2\left( -1+\frac{m^2_1-m^2_2}{2q^2}\right) B_0(0,m_1^2,m^2_1)
\ +\ m_2^2\left( -1+\frac{m^2_2-m^2_1}{2q^2}\right)\nonumber\\
&&\times\ B_0(0,m_2^2,m^2_2)
\ -\ \frac{q^2}{3}\ +\ \frac{(m^2_1-m^2_2)^2}{2q^2} \Bigg],
\eea 
where $n$ is the dimensionality of space-time and
the standard two-point integral, $B_0$, is listed, {\it e.g.},
in Appendix~A of Ref.~\cite{Bernd}.
For the evaluation of the counterterms, we also need $\Pi_V$ at $q^2=0$,
\bea
\Pi_V(0,m_1,m_2) &=& \frac{1}{16\pi^2} \Bigg[\ -2(m_1-m_2)^2
\left(\frac{2}{4-n}-\gamma_E-\ln\pi-
\frac{1}{2}\ln(m^2_1m^2_2) \right)\nonumber\\
&&-m^2_1-m^2_2+4m_1m_2
+\frac{m^4_1+m^4_2-2m_1m_2(m^2_1+m^2_2)}{m^2_1-m^2_2}\ln\frac{m^2_1}{m^2_2}
\Bigg],\nonumber\\
&&
\eea 
where $\gamma_E$ is Euler's constant.
In fact, $\Pi_V(0,m,m)=0$ as required by electromagnetic gauge
invariance.
The pseudo-scalar and axial-vector two-point functions emerge from
Eqs.~(A.1) and (A.2) by $\gamma_5$ reflection, {\it i.e.},
$\Pi_P(q^2,m_1,m_2)=-\Pi_S(q^2,m_1,-m_2)$ and
$\Pi_A(q^2,m_1,m_2)=\Pi_V(q^2,m_1,-m_2)$, respectively.
In our calculation, we have used these properties to eliminate the
$\Pi_P$ and $\Pi_A$ functions.

In our analysis, all vertex corrections can be reduced to the basic
three-point integral,
\bea
\frac{1}{16\pi^2}\int\ \frac{d^nl}{i\pi^2}
\ \mbox{tr}\left(\frac{1}{\slash{l}+\slash{p}+\slash{k}-m_3}
\ \gamma^\nu\, \frac{1}{\slash{l}+\slash{p}-m_2}
\ \gamma^\mu\, \frac{1}{\slash{l}-m_1}\right)
\nonumber\\
=\ \cA p^\mu p^\nu\ +\ \cB k^\mu k^\nu \ +\ \cC p^\mu k^\nu \ +\
\cD k^\mu p^\nu \ +\ \cE \mbox{g}^{\mu\nu}.
\eea 
As explained in Sect.~3.2, only $\cD$ and $\cE$ enter our final results.
These can be expressed in terms of $B_0$ and the standard three-point
integrals $C_0$, $C_{11}$, $C_{12}$, $C_{23}$, and $C_{24}$, viz.\
\bea
\cD(m_1,m_2,m_3) &=& \frac{1}{4\pi^2}\Big[-m_1C_0\ +\
(-m_1+m_2)C_{11}\ -\ (2m_1+m_2+m_3)C_{12}\nonumber\\
&&-\ 2(m_1+m_3)C_{23}\Big], \\
\cE(m_1,m_2,m_3)&=& \frac{1}{8\pi^2}\Bigg[
(-m_1+m_2)B_0(a,m_1^2,m_2^2)\ +\ (m_2-m_3)B_0(b,m^2_2,m^2_3)\nonumber\\
&&+\ (m_1+m_3)\Big(4C_{24}-B_0(c,m^2_1,m^2_3)\Big)\ +\
\Big[m_1(-b+m_2^2+m^2_3)\nonumber\\
&&+\ m_2(c-m^2_1-m^2_3)\ +\ m_3(-a+m^2_1+m^2_2)\
-\ 2m_1m_2m_3\Big]C_0\ \Bigg],\qquad\ \
\eea 
where we have suppressed $a=p^2,b=k^2,c=(p+k)^2$ in the argument lists
of $\cD$ and $\cE$ and it is understood that the $C$ functions are
evaluated at $(a,b,c,m_1^2,m_2^2,m_3^2)$ in the notation of Ref.~\cite{Bernd}.
It is convenient to introduce the following short-hand notations:
\bea
\overline{\cD}(m_1,m_2,m_3) &=& \cD(m_1,m_2,m_3)\ +\
\cD(m_1,-m_2,m_3)\nonumber\\
&&+\ \cD(-m_1,m_2,m_3)\ -\ \cD(m_1,m_2,-m_3),\\
\cD_\pm(m_1,m_2,m_3) &=& (m_1+m_3)\Big(
\cD(m_1,m_2,m_3)-\cD(m_1,-m_2,m_3)\Big)\nonumber\\
&&\pm\ (m_1-m_3)\Big( \cD(-m_1,m_2,m_3)+\cD(m_1,m_2,-m_3)\Big),\\
\eea 
and similarly for $\overline{\cE}$ and $\cE_\pm$.

\end{appendix}

\newpage

\newpage

\centerline{\bf\Large Figure Captions }
\bigskip\bigskip
\newcounter{fig}
\begin{list}{\bf\rm Fig. \arabic{fig}: }{\usecounter{fig}
\labelwidth1.6cm \leftmargin2.5cm \labelsep0.4cm \itemsep0ex plus0.2ex }

\item Feynman diagrams pertinent to the fermionic (a) $W^+W^-H$, (b) $ZZH$,
and $Z\gamma H$ vertex corrections in models with Majorana neutrinos.

\item Radiative corrections to the $H\to t\bar{t}$ decay width
induced by a fourth generation with Dirac neutrinos
($m_D=m_{N_1}=m_{N_2}$) as a function of $M_H$
for selected values of $m_D$
assuming $m_E=m_D$ and $m_{t'}=m_{b'}=400$~GeV.

\item Radiative corrections to the $H\to t\bar{t}$ decay width
induced by a fourth generation with Dirac neutrinos
($m_D=m_{N_1}=m_{N_2}$) as a function of $m_D$
for selected values of $M_H$
assuming $m_E=m_D$ and $m_{t'}=m_{b'}=400$~GeV.

\item Radiative corrections to the $H\to t\bar{t}$ decay width
induced by a fourth generation with Majorana neutrinos
($m_{N_1}<m_D<m_{N_2}$)
as a function of $m_{N_1}$
for selected values of $M_H$
assuming $m_D=m_E=m_{t'}=m_{b'}=400$~GeV.

\item Radiative corrections to the $H\to W^+W^-$ decay width
induced by a fourth generation with Dirac neutrinos
($m_D=m_{N_1}=m_{N_2}$) as a function of $M_H$
for selected values of $m_D$
assuming $m_E=m_D$ and $m_{t'}=m_{b'}=400$~GeV.

\item Radiative corrections to the $H\to W^+W^-$ decay width
induced by a fourth generation with Dirac neutrinos
($m_D=m_{N_1}=m_{N_2}$) as a function of $m_D$
for selected values of $M_H$
assuming $m_E=m_D$ and $m_{t'}=m_{b'}=400$~GeV.

\item Radiative corrections to the $H\to W^+W^-$ decay width
induced by a fourth generation with Majorana neutrinos
($m_{N_1}<m_D<m_{N_2}$)
as a function of $m_{N_1}$
for selected values of $M_H$
assuming $m_D=m_E=m_{t'}=m_{b'}=400$~GeV.

\item Radiative corrections to the total cross section of $e^+e^-\to ZH$
induced by a fourth generation with Dirac neutrinos
($m_D=m_{N_1}=m_{N_2}$) as a function of $M_H$
for selected values of $\sqrt s$
assuming $m_D=m_E=m_{t'}=m_{b'}=400$~GeV.

\item Radiative corrections to the total cross section of $e^+e^-\to ZH$
induced by a fourth generation with Dirac neutrinos
($m_D=m_{N_1}=m_{N_2}$) as a function of $m_D$
assuming $m_D=m_E$ and $m_{t'}=m_{b'}=400$~GeV.
The dashed (solid) line refers to low-mass (high-mass) Higgs-boson
production at LEP200 (a 500-GeV linac).

\item Radiative corrections to the total cross section of $e^+e^-\to ZH$
induced by a fourth generation with Majorana neutrinos
($m_{N_1}<m_D<m_{N_2}$)
as a function of $m_{N_1}$
assuming $m_D=m_E$ and $m_{t'}=m_{b'}=400$~GeV.
The dashed (solid) line refers to low-mass (high-mass) Higgs-boson
production at LEP200 (a 500-GeV $e^+e^-$ linac).

\end{list}

\end{document}